\documentstyle[psfig]{l-aa}

\footnotesize
\newdimen\digitwidth    
\setbox0=\hbox{\rm0}
\digitwidth=\wd0
\catcode`!=\active
\def!{\kern\digitwidth}
\normalsize

\begin{document}

 \thesaurus{11	
 	    (11.13.2; 11.08.1; 11.09.4; 11.09.1 (M31); 02.16.2)}
\title{New clues to the magnetic field structure of M31}
 \author{ J.L. Han\inst{1,2}
	 \and
	 R. Beck\inst{1}
	 \and
	 E.M. Berkhuijsen\inst{1}
         }

 \offprints{J.L. Han}

 \institute{ Max-Planck-Institut f\"ur Radioastronomie,
             Auf dem H\"ugel 69,
             D-53121 Bonn, Germany 
 \and        Beijing Astronomical Observatory,
	     Chinese Academy of Sciences,
	     Beijing 100080, China.
	     E-mail: hjl@class1.bao.ac.cn	
              }

 \date{Received 27 January 1998 / Accepted 21 April 1998}


 \maketitle

\begin{abstract}

We observed 21 polarized background radio sources in the field of M31
at 1.365 GHz and 1.652 GHz, and determined their rotation measures (RMs).
The RM data show that the regular magnetic field of the disk probably
extends from about 5 kpc to 25 kpc from the center with similar structure.
The RMs obtained from the polarized emission from M31 at $\lambda$6
cm and $\lambda$11 cm indicate that M31 might have a  weak poloidal field
in its tenuous halo. Observational features of the odd and even dynamo
modes in a galaxy are discussed. An even mode (S0) dynamo may operate
in M31.

\end{abstract}

\section{Introduction}

\begin{figure*}		
\psfig{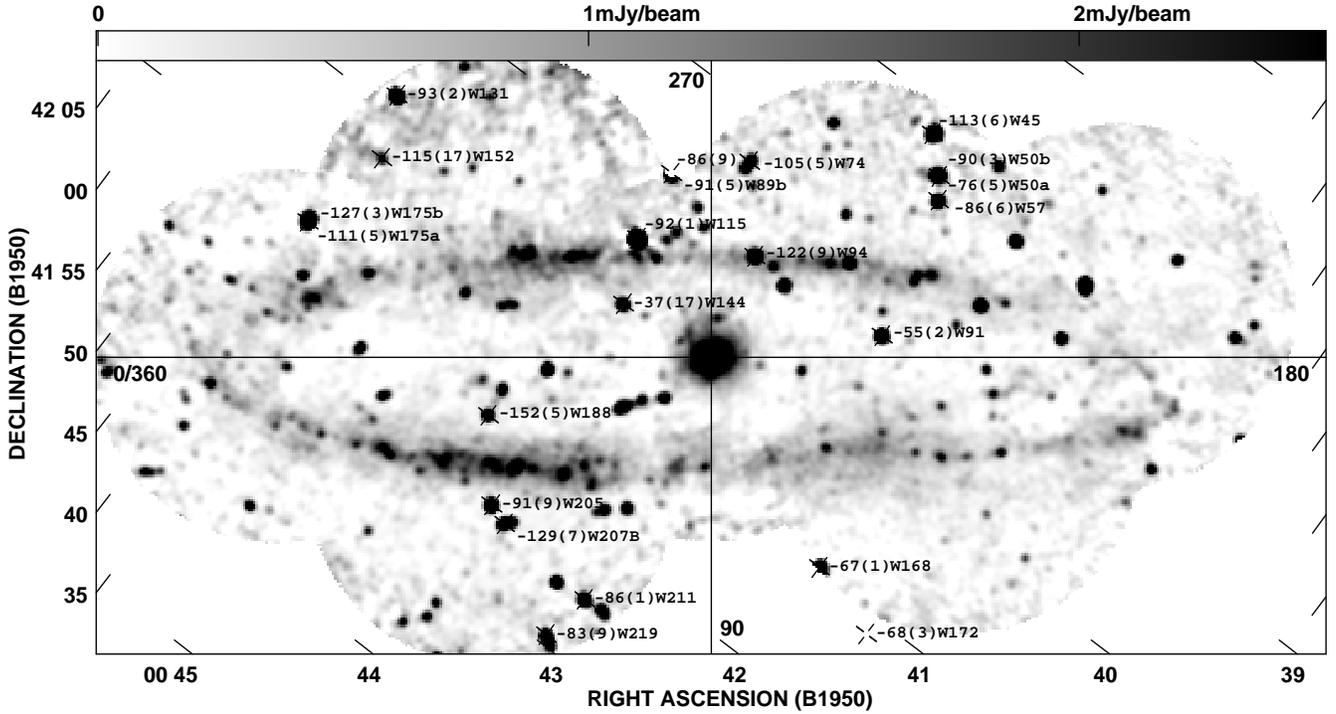}
\vspace{-5mm}
\caption{The polarized radio sources observed in this work,
marked as crosses, are superimposed onto the $\lambda20$ cm
radio emission from M31 (Beck et al. 1998). The numbers near
the sources are the RMs with their 1$\sigma$ standard deviation
in brackets, followed by their 37W catalog name. Four azimuthal
angles in the inclined galactic plane are indicated, which are 
the same as those in the sky plane by definition.}
\label{fig1}
\end{figure*}

The bright ``ring'' of nonthermal radio emission (Pooley \cite{poo69};
Beck et al. \cite{becket98}. See also Fig.~\ref{fig1}) at a radius of
about 10 kpc in the disk of M31 (NGC 224, the Andromeda Nebula) is often
referred to as basic evidence for the lowest mode of an axisymmetric
dynamo (eg. Beck et al. \cite{becket96}). The regular magnetic field
is aligned  along the ``ring'' (Beck et al. \cite{becket80}; Beck
 \cite{beck82}; Beck et al. \cite{becket89}). Outside the ``ring'',
however, very little is known about the magnetic field, mainly because
the polarized radio emission is very weak and superimposed onto extended
emission from a foreground Galactic spur (Berkhuijsen \cite{ber72};
Gr\"ave et al. \cite{graet81}).

The radio disk in galaxies often is  radially  more extended  
than the optical disk, sometimes also vertically. For example,
radially extended radio emission was detected in M51 (see Fig.1 of
Berkhuijsen et al. \cite{beret97}). Radio emission extending far away
from the galactic plane was observed from NGC 4631 (Hummel et al.
 \cite{humet91}), NGC 891 (Sukumar \& Allen \cite{sa91}) and a few
other galaxies (eg. NGC 3432 by English \& Irwin \cite{ei97}). In M31,
extended {\sc Hi} and optical emission was detected up to more than
 $0.5\degr$ from the center along the minor axis (Emerson \cite{eme74};
 Innanen et al. \cite{innet82}). However, the radio emission seems
not that extended and the existence of a radio halo in M31 is still
uncertain (eg. Wielebinski \cite{wie76};  Volodin \& Dagkesamanskii
 \cite{vd78}; Gr\"ave et al. \cite{graet81}; Berkhuijsen et al.
 \cite{beret91}). Our Galaxy has a vertically extended thick radio
disk (Beuermann et al. \cite{beuet85}) and rotation measures of
extragalactic radio sources have revealed the existence of an
extended magneto-ionic disk (Clegg et al. \cite{cleet92}; Han
 \& Qiao \cite{hq94}).  In this paper, we will investigate the
extended magneto-ionic disk in M31.

The detection of the regular magnetic field outside the ``ring'',
either interior to the ``ring'' or in the outer spiral arms or halo,
will constrain the theoretical models for the type and origin of the
field (eg. Poezd et al. \cite{poeet93};
Howard \& Kulsrud \cite{hk97}). A field residing just in the ``ring''
is very difficult to understand in the frame-work of a primordial field
origin. More seriously, a dynamo cannot generate a field that is
limited to a given small range of radius. The field should be much
more extended (Moss et al. \cite{moset98}), and should have another
weaker ring exterior (Ruzmaikin et al. \cite{ruzet88}) or interior to
the observed ``ring'' (Moss et al. \cite{moset98}). The emission possibly
detected interior to the ``ring'' from the inner arms (eg. Beck
\cite{beck82}; Berkhuijsen et al. \cite{beret91}) is too weak for a
measurement of the regular field.

M31 is a nearby spiral galaxy which optically extends more than
 $4\degr$ along the major axis on the sky. There are a number of
polarized background radio sources in the field of M31, which can
be used as probes of the magneto-ionic medium in M31. If the magnetic
structure in the halo or extended disk is somewhat ordered, the rotation
measures (RMs) of these sources should have systematic deviations from
the average. We observed 21 polarized radio sources with the VLA in 6
fields in the direction of M31 at two frequencies, and determined their
RMs from the observed position angles (PAs). We present the observations
and data reduction in Sect.2, the results in Sect.3, and discuss them
in Sect.4.
\begin{table}		
\caption{Positions of the observed fields}
\begin{tabular}{cccl}
\hline
\hline
Field & RA(1950) & Dec(1950) &
\multicolumn{1}{c}{Polarized Sources}\\
No.  & h!m~!s~ & ~!$\degr$!~$'$~!$''$ & \multicolumn{1}{c}{(37W-No.)}\\
\hline
B1. & 00~38~00.0 & +41~45~00 & 45,50,57,91\\
B2. & 00~39~30.0 & +41~10~00 & 89,94,115,144\\
B3. & 00~40~30.0 & +41~40~00 & 131,152,175\\
B4. & 00~40~40.0 & +40~40~00 & 168,172\\
B5. & 00~42~00.0 & +41~23~00 & 188\\
B6. & 00~42~30.0 & +41~00~00 & 205,207B,211,219\\
\hline
\hline
\end{tabular}
\end{table}

\begin{table*}		
\begin{minipage}{180mm}
\caption{The polarized discrete sources in the field of M31}
\begin{small}
\begin{tabular}{lccrccccrrr}
\hline
\hline
Object & RA(1950) & Dec(1950) &
\multicolumn{1}{c}{I$_{\rm 1652}$} & \multicolumn{1}{c}{PI$_{\rm 1652}$}
&  p &
 \multicolumn{1}{c}{ PA$_{\rm 1652}$}
& \multicolumn{1}{c}{PA$_{\rm 1365}$} 
&\multicolumn{1}{c}{RM}
& \multicolumn{1}{c}{PA$_{\rm intri}$}& Notes %
\\
37W-- & h!m!~s! & $\degr$!~~$'$~~~$''$ &
\multicolumn{1}{c}{(mJy)}&\multicolumn{1}{c}{(mJy)}&(\%)&
\multicolumn{1}{c}{($\degr$)} &
\multicolumn{1}{c}{($\degr$)} & \multicolumn{1}{c}{rad$\cdot$m$^{-2}$} &
\multicolumn{1}{c}{($\degr$)}\\
\hline
%
%
!45a& 00~37~14.0&40~55~18.0& 15.12$\pm$0.07 & 0.55$\pm$0.06 & !4&
 91$\pm$5! & 168$\pm$5!! &$-$113$\pm$6!& 134$\pm$10 & 1,2,3 \\
!50a& 00~37~30.2&40~52~09.1& 16.00$\pm$0.34 & 0.71$\pm$0.06 & !4&
 16$\pm$5! & 133$\pm$4!! & $-$76$\pm$5!& 163$\pm$11 & 1,2,3 \\
!50b& 00~37~29.7&40~52~35.0&  7.60$\pm$0.16 & 1.70$\pm$0.07 & 22&
 25$\pm$2!& 127$\pm$1!! & $-$90$\pm$3!&  15$\pm$7! & 1,2,3 \\
%
%
!57 & 00~37~40.7&40~50~45.6& 22.06$\pm$0.20 & 0.56$\pm$0.06 & !3&
118$\pm$5!!&  42$\pm$3! & $-$86$\pm$6!& 101$\pm$16 & 1,3   \\
!74A/B& 00~38~25.0&41~08~31.6&              &               &      &
              &                &$-$105$\pm$5!& 140$\pm$12 & 1,2,4 \\[1mm]
%
%
!89a& 00~38~55.2&41~14~04.8& 13.99$\pm$0.17 & 0.52$\pm$0.06 & !4&
 34$\pm$5! &139$\pm$5!! & $-$86$\pm$9!&  18$\pm$20 & 1,2,3 \\
!89b& 00~38~55.3&41~14~12.7& 13.45$\pm$0.17 & 1.08$\pm$0.07 & !8&
 23$\pm$3! &123$\pm$3!! & $-$91$\pm$5!&  15$\pm$11 & 1,2,3 \\
!91 & 00~38~57.2&40~47~08.0& 41.54$\pm$0.77 & 1.35$\pm$0.07 & !3&
137$\pm$2!! & 89$\pm$1! & $-$55$\pm$2!& 66$\pm$5! & 1,2,3 \\
!94a& 00~39~04.1&41~02~20.9& 18.62$\pm$0.17 & 0.54$\pm$0.06 & !3&
143$\pm$5!! & 36$\pm$13 &$-$122$\pm$9!& 14$\pm$32 & 1,5   \\
%
%
115 & 00~39~34.5&41~13~01.0&319.99$\pm$1.10 & 8.58$\pm$0.06 & !3&
 99$\pm$1! & 19$\pm$1! & $-$92$\pm$1!& 94$\pm$2! & 1,2,3 \\[1mm]
131 & 00~39~51.2&41~41~20.6& 65.05$\pm$0.56 & 1.88$\pm$0.06 & !3&
 70$\pm$2! & 168$\pm$1!! & $-$93$\pm$2!& 68$\pm$5! & 3,5   \\
144 & 00~40~07.3&41~10~09.3& 21.26$\pm$0.15 & 0.39$\pm$0.05 & !2&
 50$\pm$7! & 18$\pm$13 & $-$37$\pm$17&120$\pm$35 & 3,5   \\
152 & 00~40~23.5&41~38~43.4&  1.86$\pm$0.16 & 0.17$\pm$0.07 & !9&
 80$\pm$13 & 159$\pm$7!! &$-$115$\pm$17&118$\pm$45 & 3,5   \\
168 & 00~40~56.8&40~38~04.8& 52.59$\pm$4.53 & 2.90$\pm$0.05 & !5&
177$\pm$1!! & 118$\pm$1!! & $-$67$\pm$1!&124$\pm$1! & 3,5   \\
%
%
172 & 00~41~10.0&40~30~10.2&143.89$\pm$2.26 & 2.70$\pm$0.14 & !2&
 81$\pm$2! &  21$\pm$2! & $-$68$\pm$3!& 31$\pm$6! & 3,5   \\[1mm]
175a& 00~41~15.2&41~40~53.0& 39.69$\pm$0.35 & 1.18$\pm$0.06 & !3&
 43$\pm$4! & 123$\pm$2!! &$-$111$\pm$5!& 74$\pm$12 & 3,5   \\
175b& 00~41~13.0&41~40~52.7& 45.82$\pm$0.60 & 2.09$\pm$0.06 & !5&
 12$\pm$2! &  80$\pm$2! &$-$127$\pm$3!& 73$\pm$6! & 3,5   \\
188 & 00~41~39.6&41~14~17.7&  5.80$\pm$0.28 & 0.29$\pm$0.05 & !5&
178$\pm$4!! &  44$\pm$3! &$-$152$\pm$5!&106$\pm$13 & 3,5   \\
205 & 00~42~16.9&41~08~30.1& 31.33$\pm$1.99 & 0.81$\pm$0.06 & !3&
 90$\pm$1! &  11$\pm$13 & $-$91$\pm$9!& 83$\pm$18 & 3,5   \\
207B& 00~42~21.0&41~06~17.0&  3.72$\pm$0.33 & 0.55$\pm$0.10 & 15&
 42$\pm$3! & 110$\pm$5!! &$-$129$\pm$7!&108$\pm$15 & 3,5   \\[1mm]
211 & 00~42~27.0&40~55~09.0& 20.89$\pm$0.34 & 4.20$\pm$0.06 & 20&
 38$\pm$1! & 143$\pm$1!! & $-$86$\pm$1!& 21$\pm$2! & 3,5   \\
219 & 00~42~54.6&40~56~03.4& 11.73$\pm$0.20 & 0.50$\pm$0.05 & !4&
 42$\pm$3! & 148$\pm$7!! & $-$83$\pm$9!& 20$\pm$18 & 5     \\
\hline
\hline
\end{tabular}\\
Notes: 1. PA value at 21 cm (1490 MHz) from Beck et al. (\cite{becket89})
is also considered; 2. As note 1 but at 6.3 cm (4760 MHz); 3. PA value at
1400 MHz from NVSS catalog (Condon et al. \cite{conet98}) is considered;
4. No flux desities are given because of no detection in our
high-resolution observations. The RM is taken from Beck et al.
(\cite{becket89}); 5. PA value at 1465 MHz from Beck et al. 
(\cite{becket98}) is considered.
\end{small}
\end{minipage}
\end{table*}

\section{Observations and data reduction}

The observations were carried out at 1.365 GHz and 1.652 GHz
simultaneously on October 19th, 1995,  with the VLA B-array.
We observed consecutively each of the 6 fields (pointings) for
about 10 minutes. In all, three sets of 10 minutes of $uv$
data were obtained for each field in three hours observation
time. The coordinates of the field centers and the sources
detected in the fields are listed in Table 1.

For our purpose, a high-resolution array is necessary that is
``blind'' to the extended radio emission of M31 without resolving
the sources. Therefore, we chose the B-array with an angular
resolution of about $4''$ for our program. One observation band
of 50 MHz width was centered on 1.365 GHz and another one of
25 MHz width on 1.652 GHz. 3C48 (B0134+329) and 3C138 (B0518+165)
were observed at the beginning and the end, respectively, for
primary flux calibration and polarization calibration.

We processed the data with the AIPS software package. After
standard calibration and editing procedures, we used {\sc imagr}
with ``robust'' weighting to produce I, Q, U maps. Self-calibration
iterations in phase and amplitudes were applied. Finally, we
obtained the maps of linear polarized intensity (PI) and position
angle (PA).

\begin{figure*}	
\hspace{10mm}
\psfig{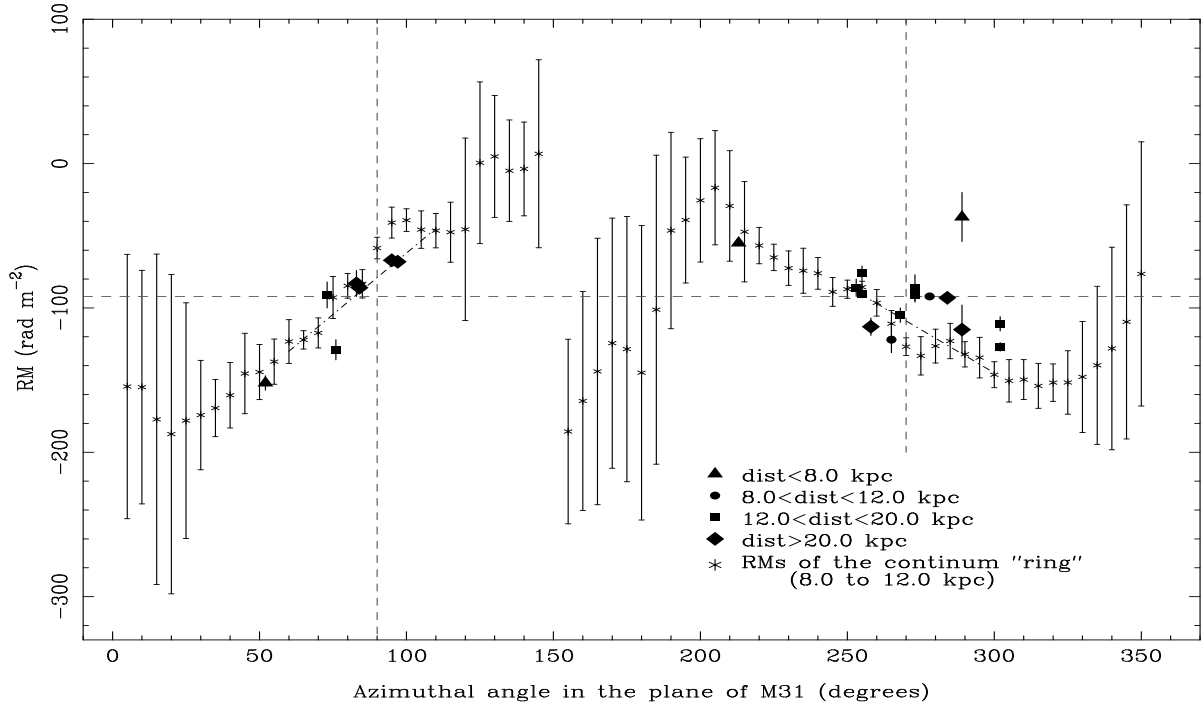}
\vspace{-2mm}
\caption{The RMs of the sources in Fig.1 and Table 2 are shown
on the RM variation along the ``ring'' deduced from Effelsberg
 $\lambda$6 cm and $\lambda$11 cm polarization observations
(Berkhuijsen et al. in preparation). For some sources the error
bar is invisibly small. The anomaly around angles
$0\degr/360\degr$ and $150\degr$ is due to a well-known disturbance
of the magnetic field in the ``ring'' (Beck 1982). The average
foreground RM (= $-93\pm3$ rad~m$^{-2}$) and the azimuthal angles
of $90\degr$ and $270\degr$ are indicated by dashed lines. 
}
\label{fig2}
\end{figure*}
\section{Rotation measures}

The sources with significant linear polarization detected are
shown in Fig.~\ref{fig1} and listed in Table 2. We give the
identification numbers of the 37W catalog (Walterbos et al.
 \cite{walet85}) and positions from our observations. The total
intensities I and polarized intensities PI at 1652 MHz are
listed in the Table. (The data of these sources at 1365 MHz
are available from the authors via email.) For some extended
sources such as 37W207B and 37W219, we give the positions of
the polarization peak. The uncertainty of the positions is
normally about $1''$. The RMs were obtained by comparing the
position angles at 1365 MHz and 1652 MHz. Published position
angles at 21 cm (1490 MHz) and 6.3 cm (4750 MHz) from Beck et
al. (\cite{becket89}), at 1400 MHz from the NVSS catalog
(Condon et al. \cite{conet98}) and at 1465 MHz from Beck et al.
(\cite{becket98}) were considered to solve the RM ambiguities,
as indicated in the notes. For some  resolved sources (eg.
37W050 with 4 components, 37W045 and 37W089 with two components),
we list each polarized component separately with a mark ``a'' or
``b'' following its 37W catalog name. The RM of 37W074A/B from
Beck et al. (\cite{becket89}) is included for the following
discussion.

There are four or five possible contributions to the observed
RMs of these sources: the foreground RM from our Galaxy, the
RM contribution from the disk and/or the halo of M31, the
intergalactic RM contributions (between M31 and the sources
and between  M31 and our Galaxy), and the intrinsic RMs from
the sources themselves. The RM from the intergalactic medium
is believed to be quite small. The intrinsic RMs of sources
are difficult to address. The sum of the last two contributions
is usually smaller than 10 rad~m$^{-2}$ for most sources
as justified by the small RMs of sources at the Galactic poles
(Simard-Normandin \& Kronberg \cite{sk80}; Han et al. 1998),
and therefore will not be considered below. The RM sky of
our Galaxy is correlated in areas of about $10 \sim 15\degr$
radius at the Galactic latitude of M31 (Oren \& Wolfe \cite{ow95};
Simard-Normandin \& Kronberg \cite{sk80}). The Galactic spur in front
of M31 is also larger than M31 (Gr\"ave et al. \cite{graet81}). 
In the region of sky $2\degr$ around the direction to M31
 $[(l, b) =(121.2, -21.6)]$, the fluctuations of foreground
RM from our Galaxy should be small enough to ignore. The average
foreground RM from our Galaxy over the observed area is $-93\pm3$
rad~m$^{-2}$ (Berkhuijsen et al. in preparation).

Inspection of Fig.~\ref{fig1} shows that most RM values of
the background sources do not deviate much from the average
foreground RM. The RMs of three sources interior to the continuum
``ring'' differ most strongly. The RMs of most sources, however,
are more or less consistent with the RM variation along the
``ring'' shown in Fig.~\ref{fig2} (see Sect.4.1). 

\section{Discussion}
\begin{figure*}
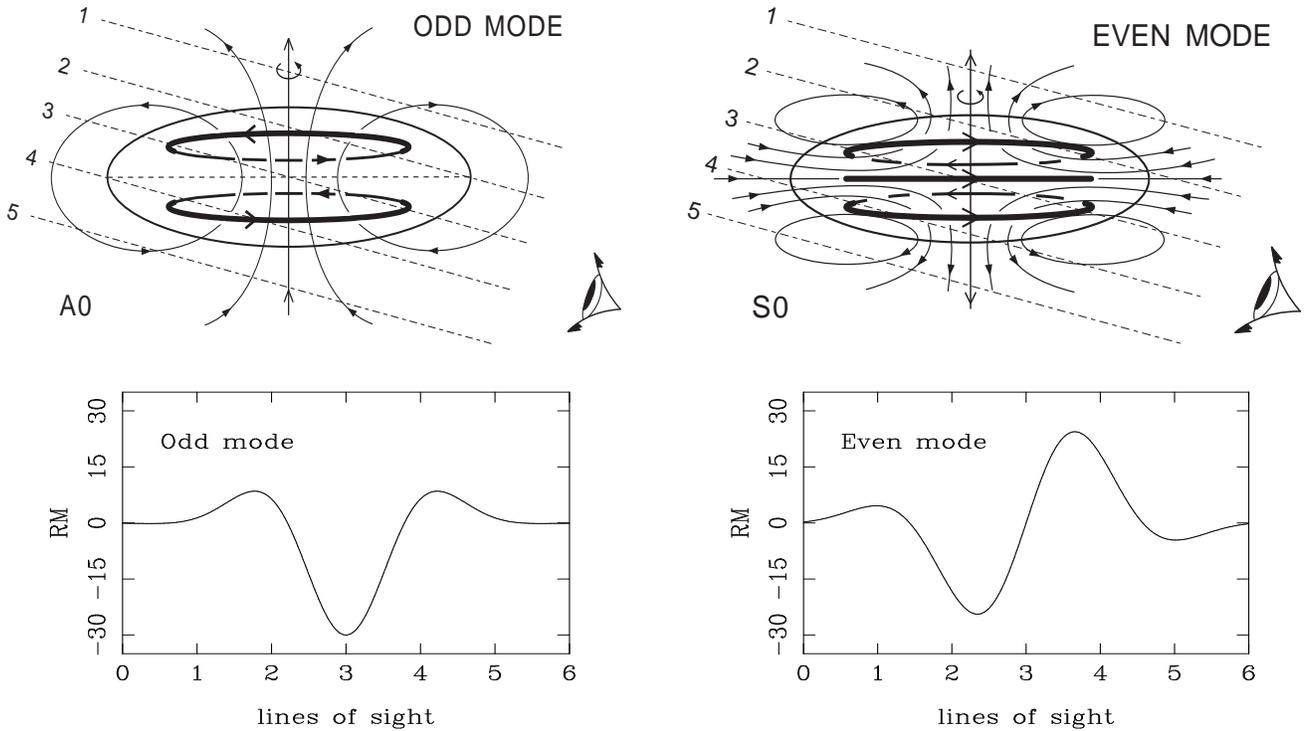
	
\centering
\begin{tabular}{ccc}
\mbox{\psfig{file=0867.f3a,height=45mm,width=82mm}} & &
\mbox{\psfig{file=0867.f3b,height=45mm,width=82mm}} 
\\[5mm]
\mbox{\psfig{file=0867.f3c,rotate=90,height=45mm,width=70mm}} & &
\mbox{\psfig{file=0867.f3d,rotate=90,height=45mm,width=70mm}} 
\end{tabular}\\[-1mm]
\caption{The field configurations of the odd (A0) and the even
(S0) dynamo modes in a galaxy, with thick lines indicating the toroidal
field and thin lines the poloidal field, are illustrated in the
two plots above; the RM of polarized background sources from the
poloidal field is shown in the two plots below (with arbitrary scale
for RM). The background radio sources are supposed to be located in
the plane of the azimuthal angles of $90\degr$ and $270\degr$ (the
minor axis). The RM of the extended polarized emission from a galaxy,
eg. from the ``ring'' of M31, near lines of sight No.1 and No.2 have
smaller amplitudes than those shown in the lower plots. 
}
\label{fig3}
\end{figure*}

If the magnetic field in a galaxy is generated or maintained
by a dynamo, theoretical simulations show that the lowest mode
dynamo is excited most easily (eg. Beck et al. \cite{becket96}).
The two lowest modes, the odd mode A0 and the even mode S0,
could exist in galaxies. Observational proofs of these modes
are obviously important, not only to further theoretical studies
but also to discrimination between the dynamo theory and its
alternatives, eg. primordial origin (eg. Kulsrud et al.
\cite{kulet97}), local origin (eg. Daly \& Loeb \cite{dl90}),
or MHD waves (eg. Fan \& Lou \cite{fl96}). 

As illustrated by Fig.19 in Wielebinski \& Krause (1993), the
toroidal field in the A0 mode is antisymmetric with respect to
the galactic plane, and the poloidal field has the structure of
a dipole (see our Fig.~\ref{fig3}). The S0 mode field, however,
has the same toroidal field structure above and below the plane,
but an antisymmetric poloidal field, like two opposite dipoles. 

Our Galaxy is an edge-on galaxy. Han et al. (\cite{hanet97})
analysed the RM distribution of several hundred background
sources, and showed that our Galaxy possibly has an A0 odd
mode field, eg. an antisymmetric toroidal field with respect
to the Galactic plane. For M31, the well-known ``ring'' is
thought to be an indication of a toroidal field, but it is
not clear yet whether it is of odd or even mode.

Polarization observations of M31's extended emission show
that there is a systematic RM variation around the observed
``ring'' located at a galacto-centric radius of about 10 kpc,
as shown in Fig.~\ref{fig2}, which is caused by the regular
magnetic field inside the ``ring'' (Beck \cite{beck82};
Berkhuijsen et al. \cite{beret87}; Beck et al. \cite{becket89};
Berkhuijsen et al. in preparation). The gas layer causing the
observed RM variations is primarily the near half, in contrast
to the full thickness of the disk where the emission comes
from. The RM variation with azimuthal angle of such a continuum
``ring'' should have the same shape for magnetic fields of
the A0 odd mode and of the S0 even mode, but they differ in
amplitude by a factor of 2 according to Beck et al. (\cite{becket96}).
This difference cannot be used as a criterion to distinguish A0 or S0
mode, because it is not known to which mode the observed amplitude
of the RM curve refers.

The RMs of the background sources reveal the average magnetic
field across the full thickness of the disk. If the toroidal
magnetic field of an S0 even dynamo mode is uniform in the disk
of a galaxy, the RMs of background sources should be twice that
of the extended emission from M31. On the other hand, for an A0
odd dynamo field the RM contribution from the toroidal field in
the near half and the far half of the disk of M31 will cancel
each other and therefore there will be no net RM contribution
when observing background sources. 

Observations of the RMs near the minor axis, either
of background sources or of the extended emission from the ``ring'',
could provide evidence for a symmetric poloidal field of an even
dynamo mode or an antisymmetric poloidal field of an odd dynamo
mode. The toroidal field near these azimuthal regions is almost
perpendicular to the line of sight and therefore contributes a
negligible RM. Because of the poloidal field,  the RMs of the
sources along the lines of sight No.1 and No.5 (or No.2 and No.4)
should similarly deviate from the average foreground RM in the odd
mode, while opposite deviations should be found in the even mode
(Fig.~\ref{fig3}).

In the following we will consider that the deviating RMs of the
observed sources are caused by the magneto-ionic medium in either
the disk or the halo of M31, or both.
\begin{table}	
\caption{The geometric parameters of polarized sources in order
of $\theta_{\rm sky}$}
\begin{tabular}{lccccc}
\hline
\hline
Object   & RM &  $\theta_{\rm sky}$ $^1$ & $\theta_{\rm M31}$ $^{1,2}$  & 
$R_{\rm M31}$ $^2$\\
37W--   & (rad m$^{-2}$) & ($\degr$) & ($\degr$)  & (kpc) \\
\hline
188 & $-152\pm5!$ & !15 & !52 & !7.3  \\
205 & $!-91\pm9!$ & !34 & !73 & 15.1  \\
207B& $-129\pm7!$ & !39 & !76 & 16.8  \\
219 & $!-83\pm9!$ & !59 & !83 & 27.1  \\
211 & $!-86\pm1!$ & !62 & !84 & 23.7  \\
168 & $!-67\pm1!$ & 116 & !96 & 20.6  \\
172 & $!-68\pm3!$ & 118 & !96 & 26.0  \\
!91 & $!-55\pm2!$ & 188 & 213 & !4.9  \\
!57 & $!-86\pm6!$ & 215 & 253 & 15.8  \\
!50 & $!-76\pm5!$ & 219 & 255 & 18.3  \\
    & $!-90\pm3!$ &     &     &       \\
!45 & $-113\pm6!$ & 225 & 258 & 22.3  \\
!94 & $-122\pm9!$ & 246 & 265 & 10.0  \\
!74 & $-105\pm5!$ & 259 & 268 & 18.7  \\
!89 & $!-86\pm9!$ & 282 & 273 & 17.6  \\
    & $!-91\pm5!$ &     &     &       \\
115 & $!-92\pm1!$ & 302 & 278 & 11.4  \\
131 & $!-93\pm2!$ & 321 & 284 & 26.4  \\
144 & $!-37\pm17$ & 329 & 289 & !5.3  \\
152 & $-115\pm17$ & 329 & 289 & 20.3  \\
175 & $-127\pm3!$ & 342 & 302 & 15.4  \\
    & $-111\pm5!$ &     &     &       \\
\hline
\hline
\end{tabular}\\
Notes: (1) Azimuthal angles $\theta = 0\degr$, $90\degr$, 
 $180\degr$ and $270\degr$ are shown in Fig.1. Just for
these values, they are the same in the plane of sky and
in the plane of M31. (2) We assumed an inclination of
 $78\degr$ and a distance to M31 of 690 kpc.
\end{table}

\subsection{On the toroidal field}

The toroidal magnetic field in M31 results in a systematic RM
variation around the observed ``ring'' (Fig.2). The slight
phase shift of the RM curve (RM$\sim 0$ at $\theta_{\rm M31}
 \simeq  80\degr$ and $250\degr$) is due to the pitch angle
of the field in the ``ring''. A detailed discussion of this
RM variation, including the big jumps (Beck 1982) near
 $\theta_{\rm M31} = 150\degr$ and $0\degr$/$360\degr$,
will be given by Berkhuijsen et al. (in preparation). 

The field in the ``ring''  contributes to RMs of the discrete
sources if they are located behind the ``ring''. However, if
a source has a large angular separation from the ``ring'',
then the field in the ``ring'' will not affect its RM unless
the field extends to that large radius. In Table 3, we give
the apparent radial distances $R_{\rm M31}$ of the sources
from the M31 center and their azimuthal angle $\theta_{\rm M31}$,
both in the galactic plane of M31. These two parameters indicate
which part of M31 affects the observed RM of the background sources.
With an inclination angle of $78\degr$ (eg. Braun \cite{bra91};
Ma et al. \cite{maet97}) and a distance\footnote{
Recently, Feast \& Catchpole (\cite{fc97}) published the M31
modulus $24.77\pm0.11$, implying a larger distance to M31 of
$900\pm45$ kpc. For comparison with earlier work, we use 690
kpc.}
to M31 of 690 kpc (Baade \& Swope \cite{bs63}), $1'$ corresponds
to 200 pc in the plane of M31 along the major axis. The azimuthal
angle increases counter-clockwise from $0\degr$ on the northern
major axis to east (see Fig.~\ref{fig1}). We plotted the observed
RMs of discrete sources onto Fig.~\ref{fig2} with different symbols
indicating the range of apparent distances to the center of M31.

Most of the observed sources are near the azimuthal angles
 $\theta_{\rm M31} = 90\degr$ and $270\degr$ (Fig.~\ref{fig2}),
and therefore have a RM not deviating much from the average RM or
from the RM variation of the ``ring'', except for a few.

Three sources interior to the ``ring'' (see Fig.~\ref{fig1}) with
quite different azimuthal angles $\theta_{\rm M31}$ have apparent
distances less than 8.0 kpc (Tab.2) and are located far from the
minor axis of M31. Note that two of them, 37W188 (RM =$-152\pm5$)
and 37W91 (RM =$-55\pm2$), have RMs quite consistent with those
of the continuum ``ring''. First of all, this indicates that the
field with the same structure as in the ``ring'' also extends to
the inner part of M31. Second, the field very probably has an
even mode, eg. no field reversal below and above the galactic
plane of M31. Otherwise there should be no net RM from M31. Third,
this means that the strength of the regular magnetic field has
about one half of the strength compared to that in the ``ring'',
if the electron density were constant. The third source, 37W144
(RM =$-37\pm17$), has a large positive RM after the foreground RM
is subtracted, opposite to that of the continuum ``ring''. This
source is exactly located behind a dust lane
(Walterbos \& Kennicutt \cite{wk88}) and an {\sc Hi} spiral arm
(Brinks \& Shane \cite{bs84}). Comparison with the spiral structure
of M31 (Braun \cite{bra91}) suggests some strong local perturbation
of the magnetic field from the {\sc Hii} regions in this direction,
or possibly a field reversal in the dust arm.

Two sources appear on the ring, 37W94 (RM =$-122\pm9$) and 37W115
(RM =$-92\pm1$). Both of them probably are background sources, as
indicated by the {\sc Hi} absorption lines (Braun \& Walterbos
\cite{bw92}). The former one has an RM deviating from the average
twice as much as the continuum emission, though with large error
bars, indicating that the field there might be of an even mode as
discussed above. 37W115 is the strongest source detected. This
source has soft X-ray emission (Supper et al. \cite{supet97}),
and its X-ray hardness (Supper et al. \cite{supet97}) suggests
that it is a background source. However, there seems to be no
net RM contribution from M31 so that the RM of this source is
equal to the average (eg. the foreground RM). This could happen
if the 3 spiral arms (Braun \cite{bra91}), through which the
line of sight to 37W115 passes, have reversed field directions.
The RM of the extended polarized emission from the ``ring''
is not cancelled because it mainly emerges from one arm (Berkhuijsen
 et al. \cite{beret93}). Another sources, 37W89 (RM =$-86\pm9$ and
 $-91\pm5$), not on the ring, is located at a similar
$\theta_{\rm M31}$, but this source may be a supernova remnant
within M31 (Dickey \& Brinks \cite{db88}), which explains the
small RM contribution from M31.

Sources exterior to the ``ring''  are interesting as well. Compared
with the spiral structure given by Braun (\cite{bra91}), we found
that only 37W131 (RM =$-93\pm2$) is far away from the spiral arms,
and has an RM equal to the average, which means that the magneto-ionic
medium does not extend to $R_{\rm M31} = 26.4$ kpc in this part of
M31. However, in the lower part of Fig.1, most sources are in the
spiral region outlined by Braun (\cite{bra91}) and their RMs are
consistent with the RM variation of the ``ring''. This suggests
that the magnetic field in the disk extends as far as
 $R_{\rm M31} = 26$kpc in this region, as indicated by 37W168
(RM =$-67\pm1$) and 37W172 (RM =$-68\pm3$). Furthermore, the
toroidal field at this radius has the same direction as that
in the ``ring'', and a significant strength.

The RMs of 37W175 (=$-127\pm3$ and $-111\pm5$) suggest that the
field with the same structure as in the ``ring'' extends to
$R_{\rm M31} = 15$ kpc but with about one half or one third of
the field strength in the ``ring'', assuming constant electron
density. Three sources at the upper right corner of Fig.1 have
about average RMs, similar to the RM of continuum ``ring'' at
these azimuthal angles. Only the farthest one, 37W45 (RM
=$-113\pm6$), has a somewhat negative deviation,
which suggests that a weak field extends even to $R_{\rm M31} = 22$
kpc. This also holds for 37W74 (RM =$-105\pm5$). 

In summary, our results suggest that (1) magnetic field exists
exterior  as well as interior to the ``ring''; (2) the magneto-ionic
disk of M31 is extended, probably
out to 25 kpc in the plane of the disk; and (3) the field in the
extended disk generally has a  well-ordered structure similar to
that in the ``ring'', but the field strength is smaller than in
the ``ring'', in agreement with the very weak synchrotron emission
observed from the extended disk (Berkhuijsen \& Wielebinski \cite{bw74}).

\subsection{On the poloidal field}

The RM data of the continuum ``ring'' near the azimuthal angles
of $90\degr$ and $270\degr$ seem to give some indication for the
poloidal magnetic field (in the halo and disk). As is shown in
Fig.3, the RMs along lines of sight No.2 and No.4 are expected
to have the largest opposite deviations from the mean curve
caused by RMs originating in the poloidal field which are added
to those of the toroidal field. Indeed, two opposite RM bumps,
though with  large error bars, appear in Fig.2 on the RM variation
of the ``ring'' near $\theta_{\rm M31} = 100\degr$ and $280\degr$,
which could be  caused by the poloidal field (Fig.3). Since the
part of M31 at $\theta =270\degr$ is the near one, we
see a smaller bump. This indicates that the poloidal field of
M31 is of even mode as is shown for the S0 dynamo field in Fig.3.
The field is directed inwards near the galactic plane, but is
oppositely going outwards when it is far from plane. If this
opposite deviation could be confirmed by more RM data of
background sources, then the magnetic field in the halo of M31 is
most probably generated by an even mode dynamo,  consistent with
the ring field in the disk. The toroidal field in the M31 disk
should then be mirror-symmetric with respect to its galactic plane
without a vertical field reversal.

No matter what kind of dynamo operates in the halo of M31, if we
attribute the deviating RMs of the bumps, $\Delta RM \sim 15$
rad~m$^{-2}$, to the poloidal field, we can estimate an upper
limit of the field strength. Assuming a thermal electron density
of  0.003 cm$^{-3}$ and a line of sight of  20 kpc through the
thick disk or halo, then this upper limit is about 0.3$\mu$G,
almost the same as the halo field of our Galaxy near the sun
(Han \& Qiao 1994; Han et al. \cite{hanet98}).

\subsection{Concluding remarks}

Our RM observations of background radio sources suggest that
the regular magnetic field in the bright ``ring'' of M31 extends
from a radius of about 5 kpc interior to the ``ring'' to as far as
25 kpc from the center, probably with  similar structure.
We presented evidence that the magnetic field of M31 has the
structure of an even mode dynamo field (S0). The poloidal
field of even symmetry has a strength $\le0.3\mu$G.

\acknowledgements 
We are grateful to Prof. P.P. Kronberg and Prof. R.~Wielebinski for
suggesting the project. We thank our internal referee, Prof. P.P.
Kronberg, and the referee, Dr. R.A.M. Walterbos, for their careful
reading of the manuscript and helpful comments, and Prof. M. Urbanik for
checking Fig.~3 by means of computer simulations. JLH is grateful
to Prof. R.~Wielebinski for his continuous support and stimulating
discussions. This work has been done under the exchange program
between the Chinese Academy of Sciences and the Max-Planck-Gesellschaft.
JLH acknowledges support from the Astronomical Committee of the Chinese
Academy of Sciences and the Director Foundation of BAO. The National
Radio Astronomy Observatory is a facility of the National Science
Foundation operated under cooperative agreement by Associated
Universities, Inc.

\end{document}